\documentclass[sigconf,natbib=false]{acmart}
\AtBeginDocument{%
  }

\copyrightyear{2026}
\acmYear{2026}
\setcopyright{cc}
\setcctype{by}
\acmConference[SAC '26]{The 41st ACM/SIGAPP Symposium on Applied Computing}{March 23--27, 2026}{Thessaloniki, Greece}
\acmBooktitle{The 41st ACM/SIGAPP Symposium on Applied Computing (SAC '26), March 23--27, 2026, Thessaloniki, Greece}
\acmPrice{}
\acmDOI{10.1145/3748522.3779903}
\acmISBN{979-8-4007-2294-3/2026/03}

\RequirePackage[
  datamodel=acmdatamodel,
  style=acmnumeric,
  ]{biblatex}

\usepackage{enumitem}
\usepackage{booktabs}
\usepackage{multirow}
\usepackage{adjustbox}
\usepackage{graphicx}   
\usepackage{subcaption} 
\begin{document}

\title{Early-Stage Product Line Validation Using LLMs: A Study on Semi-Formal Blueprint Analysis}
\renewcommand{\shorttitle}{Early-Stage Product Line Validation Using LLMs: A Study on Semi-Formal Blueprint Analysis}

\author{Viet-Man Le}
\email{v.m.le@tugraz.at}
\orcid{0000-0001-5778-975X}
\affiliation{%
  \institution{Graz University of Technology}
  \city{Graz}
  \country{Austria}
}

\author{Thi Ngoc Trang Tran}
\email{trang.tran@tugraz.at}
\orcid{0000-0002-3550-8352}
\affiliation{%
  \institution{Graz University of Technology}
  \city{Graz}
  \country{Austria}
}

\author{Sebastian Lubos}
\email{sebastian.lubos@tugraz.at}
\orcid{0000-0002-5024-3786}
\affiliation{%
  \institution{Graz University of Technology}
  \city{Graz}
  \country{Austria}
}

\author{Alexander Felfernig}
\email{alexander.felfernig@tugraz.at}
\orcid{0000-0003-0108-3146}
\affiliation{%
  \institution{Graz University of Technology}
  \city{Graz}
  \country{Austria}
}

\author{Damian Garber}
\email{damian.garber@tugraz.at}
\orcid{0009-0005-0993-0911}
\affiliation{%
  \institution{Graz University of Technology}
  \city{Graz}
  \country{Austria}
}

\renewcommand{\shortauthors}{V.M. Le et al.}


\begin{abstract}
We study whether Large Language Models (LLMs) can perform \emph{feature model analysis operations} (AOs) directly on \emph{semi-formal textual blueprints}, i.e., concise constrained-language descriptions of feature hierarchies and constraints, enabling early validation in Software Product Line scoping. Using 12 state-of-the-art LLMs and 16 standard AOs, we compare their outputs against the solver-based oracle FLAMA. Results show that \emph{reasoning-optimized models} (e.g., Grok 4 Fast Reasoning, Gemini 2.5 Pro) achieve 88--89\% average accuracy across all evaluated blueprints and operations, approaching solver correctness. We identify systematic errors in structural parsing and constraint reasoning, and highlight accuracy–cost trade-offs that inform model selection. These findings position LLMs as lightweight assistants for early variability validation.
\end{abstract}

%
%



\maketitle

\section{Introduction}

Software Product Line Engineering (SPLE) systematically manages variability and enables large-scale reuse in software-intensive systems~\cite{Apeletal2013,Clements2002}. A critical early stage is \emph{Software Product Line (SPL) scoping}, where engineers and domain experts define the boundaries of the product line, identify candidate features, and make assumptions about variability~\cite{marchezan2022}. Scoping outcomes have a profound influence on all subsequent phases, since they determine which requirements, architectures, and reusable assets will be engineered. Empirical studies confirm that incorrect or incomplete scoping decisions can cascade into costly rework and misalignment with business goals. Industrial practitioners continue to view scoping as one of the most persistent challenges in SPLE adoption~\cite{Becker2024}.

To ensure correct variability decisions, researchers have developed a broad catalogue of automated \emph{analysis operations} (AOs) for feature models (FMs)~\cite{BeSeRu2010,Galindo2019}. These operations extract information from FMs and verify properties that are difficult or infeasible to assess manually, especially as models scale to hundreds or thousands of features~\cite{BeSeRu2010}. Typical examples include detecting dead or false optional features, validating satisfiability, computing valid configurations, and estimating configuration-space size. Such analyses are vital for maintaining FM correctness and scalability~\cite{Galindo2019}. They are typically applied only \emph{after} an FM has been constructed during \emph{domain requirements engineering}, delaying feedback until late in the process when corrections are costly and disruptive~\cite{Ghosh2016,Khor2024}.

This paper investigates the potential of applying automated analysis already during the \emph{scoping} phase, where assumptions about variability are first articulated. We introduce an \emph{early validation workflow} in which these assumptions are expressed in semi-formal textual \emph{blueprints}~\cite{michailidis2024} and analyzed directly with Large Language Models (LLMs). The workflow provides immediate feedback on feasibility, consistency, and potential defects before formal feature models are constructed, thereby bridging informal scoping practices with automated validation and enabling domain experts to iteratively refine scope decisions. This early validation step avoids the need for a solver or formal knowledge base, reducing modeling effort and enabling earlier detection and discussion of design issues.


Recent advances in LLMs suggest that this approach is feasible. Studies show that, when carefully prompted, LLMs can perform deductive and logical inference beyond surface-level text understanding~\cite{huang2023,lin2025zebralogic,pan2025can,Parmar2024,Yan2024}, and can map natural language into formal specifications such as answer set programs~\cite{Ishay2023}, constraint satisfaction problems (CSPs)~\cite{Hotz2024,michailidis2024}, or Universal Variability Language (UVL) feature models~\cite{Galindo2023}. These findings indicate that LLMs may function as lightweight inference engines over semi-formal representations like blueprints. However, their out-of-the-box ability to perform feature model AOs on such inputs has not been systematically studied. This paper addresses this gap through a large-scale empirical evaluation of 12 state-of-the-art LLMs on 16 AOs, comparing their accuracy, runtime, and failure modes against a solver-based oracle.  

The contributions of this paper are threefold.
\textit{Firstly}, we formalize and motivate a workflow for \emph{early product-line validation}, which combines semi-formal blueprints with lightweight LLM-based AOs to provide feedback already during SPL scoping. \textit{Secondly}, we conduct a large-scale systematic evaluation of off-the-shelf LLMs on feature model analysis tasks using blueprint inputs. Our study covers 12 LLMs (both general-purpose and reasoning-optimized) and 16 AOs across solver-free and solver-based categories. \textit{Finally}, we systematically evaluate LLM accuracy, cost, and failure modes against the solver-based oracle FLAMA~\cite{Galindo2023flama}, providing insights into their suitability as lightweight assistants for SPL scoping.



\section{Background and Related Work}
\label{sec:background}

\subsection{Software Product Line Engineering}

The primary objective of SPLE is to reduce development cost and time-to-market while improving quality by deriving families of related products from shared assets rather than engineering each product independently \cite{Apeletal2013,Clements2002}. Its lifecycle is commonly organized into two complementary processes: \emph{domain engineering}, which establishes reusable core assets, and \emph{application engineering}, which derives concrete products from them \cite{pohl2005software}. Domain engineering is typically structured into four phases: (i) \emph{scoping}, where product-line boundaries, candidate features, and variability assumptions are defined; (ii) \emph{domain requirements engineering}, which captures common and variable requirements; (iii) \emph{domain design and implementation}, which produce reusable architectures and components; and (iv) \emph{domain testing}, which validates the shared assets before reuse \cite{Apeletal2013}. Application engineering then configures and assembles products from these assets, while evolution and maintenance activities ensure long-term adaptation of the product line \cite{Berger2020}.

\subsection{Feature Models and Analysis Operations}

A central artifact of \emph{domain requirements engineering} is the \emph{feature model} (FM), which represents a product line’s commonalities and variabilities through a hierarchical feature tree and cross-tree constraints such as \emph{requires} and \emph{excludes}~\cite{Apeletal2013,Kang1990}.
To validate FMs, the community has established automated analysis operations (AOs)~\cite{BeSeRu2010,Galindo2019}. \emph{Solver-free} AOs compute structural metrics (e.g., feature counts, tree depth), while \emph{solver-based} AOs employ SAT/CSP/BDD engines to verify semantic properties such as satisfiability, dead features, and valid configurations~\cite{Benavidesetal2013,Galindo2019}. These operations transform FMs from descriptive artifacts into analyzable models, enabling systematic detection of defects difficult to assess manually at scale~\cite{BeSeRu2010}.

The Universal Variability Language (UVL) has emerged as a unified textual DSL for feature modeling~\cite{Benavides2025}. In this study, we use FLAMA~\cite{Galindo2023flama} to execute solver-based AOs on UVL inputs as our ground-truth oracle. We focus on the Boolean level of UVL, capturing feature hierarchies and cross-tree constraints.

\subsection{SPL Scoping}

Scoping defines the boundaries of a product line, identifies candidate features, and establishes variability assumptions~\cite{marchezan2022,pohl2005software}. These decisions shape all subsequent engineering activities and are widely recognized as among the most critical yet challenging in SPLE~\cite{Becker2024}.

Current validation practices rely heavily on informal stakeholder workshops, product roadmaps, and expert judgment, leaving consistency and feasibility unchecked until formal feature models exist~\cite{Ghosh2016,Khor2024}. This delay causes late-stage rework when variability conflicts or infeasible assumptions surface during requirements engineering or implementation. Integrating automated analysis operations (AOs) directly into scoping can close this gap by providing early feedback on variability properties before formalization, enabling domain experts to iteratively refine scope decisions with confidence~\cite{BeSeRu2010}.


\subsection{Large Language Models for Reasoning}

Large Language Models (LLMs) such as GPT, Claude, and Gemini are transformer-based architectures with up to hundreds of billions of parameters, pre-trained on large text corpora~\cite{Brown2020}. Beyond traditional NLP tasks, they exhibit emerging \emph{reasoning} capabilities when properly prompted, including deductive inference, logical implication, and structured problem solving~\cite{huang2023,pan2025can,Yan2024}. Benchmarks such as LogicBench~\cite{Parmar2024} and ZebraLogic~\cite{lin2025zebralogic} confirm this potential but also reveal limitations, e.g., accuracy drops under multi-step inference or complex constraints. Studies further note that many models behave as ``greedy reasoners,'' favoring short reasoning paths and failing under negation or deep logic.


Prompting techniques, such as \emph{few-shot}, \emph{chain-of-thought}~\cite{Brown2020,Kojima2022,Wei2022}, and \emph{rationale decomposition}~\cite{huang2023}, can elicit more structured reasoning without fine-tuning. These approaches have enabled LLMs to generate formal artifacts such as answer-set programs, CSPs, and UVL feature models directly from natural language~\cite{Galindo2023,Hotz2024,Ishay2023,michailidis2024}. For product line engineering, this suggests that well-prompted LLMs may serve as lightweight inference engines for analyzing semi-formal \emph{blueprints}, providing early validation without relying exclusively on solver technology.


\section{Early Validation Workflow for SPL Scoping}
\label{sec:workflow}

We propose an \emph{early validation workflow} that integrates automated analysis directly into the scoping phase. This is achieved by introducing a lightweight, semi-formal \emph{blueprint}~\cite{michailidis2024} and leveraging LLMs to analyze it. The workflow proceeds in three steps:


\begin{enumerate}[leftmargin=*,itemsep=2pt]

  \item \textbf{Step 1 - Blueprint creation:} Domain experts consolidate scope boundaries, candidate features, and variability assumptions into a semi-formal \emph{blueprint}. A blueprint is a \emph{set of textual semi-formal constraints} that collectively specify the feature hierarchy and cross-tree relations. Examples include ``\emph{Feature A requires Feature B}'', ``\emph{Feature C can be Feature D or Feature E}'', ``\emph{Feature B excludes Feature D}'', or ``\emph{Feature E can be Feature G, Feature H, or both}''. This representation is easy to author yet structured enough for automated checks, bridging free-text scoping notes and formal UVL models. 
  \textit{Figure~\ref{fig:prompting-pipeline}} (top-left panel) illustrates a concrete \emph{blueprint} for a \textit{smartwatch product line}.


  \item \textbf{Step 2 - LLM-based analysis.} The blueprint is fed to an \emph{LLM analysis engine} that executes AOs. Each AO is guided by a tailored prompt that encodes its reasoning task, ensuring the LLM applies an appropriate inference strategy. The set of AOs is extensible, allowing any operation expressible in natural language to be incorporated into the workflow. 

  \item \textbf{Step 3 - Feedback and refinement:} The engine returns actionable findings such as inconsistencies, dead features, or product counts. Based on this feedback, the domain expert revises the blueprint and, where necessary, updates scoping outputs. All these three steps are repeated until stabilization.
\end{enumerate}

The iteration yields a \emph{validated blueprint} and updated scoping dossier. These can (i) be compiled to a formal feature model for solver-based verification and (ii) seed draft UVL generation to reduce modeling effort~\cite{Galindo2023}. In short, blueprints capture early variability decisions in a form LLMs can analyze, providing timely feedback before full formalization.



The workflow provides the context for our empirical evaluation. Our study does not assess the workflow itself but investigates a critical question it raises: \emph{Are off-the-shelf LLMs capable of performing feature model analysis operations directly on semi-formal blueprints?} The next section presents our methodology.

\section{Methodology}
\label{sec:methodology}

\subsection{Research Questions}

We evaluate whether off-the-shelf LLMs can execute AOs on semi-formal blueprints and how they compare to a solver oracle (FLAMA \cite{Galindo2023flama}). The evaluation is structured around three research questions:

\begin{enumerate}[label=\textbf{RQ\arabic*},leftmargin=*,itemsep=2pt]
  \item \textbf{Accuracy.} How accurately can different LLMs perform solver-free and solver-based AOs on blueprint inputs?

  \item \textbf{Cost.} How do end-to-end costs (runtime and token usage) relate to achieved accuracy across models?

  \item \textbf{Failure modes.} What types of errors do LLMs exhibit when performing AOs? How do these vary across model families?
\end{enumerate}

\subsection{LLMs Under Study}

We evaluate 12 publicly available LLMs, spanning both \emph{general-purpose} (e.g., GPT-4.1, Claude Sonnet~4, DeepSeek Chat) and \emph{reasoning-optimized} variants (e.g., Grok~4 Fast Reasoning, GPT-5~mini, Gemini~2.5~Pro). All models are accessed via their official APIs without any fine-tuning, enabling a direct comparison between mainstream and reasoning-focused architectures.


To ensure comparability, every model is queried under identical conditions: \texttt{temperature=0}, unrestricted context and output lengths within provider limits, and a uniform prompting and evaluation pipeline. This setup yields deterministic completions that expose each model’s inherent reasoning behavior. \textit{Table~\ref{tab:llms}} lists the evaluated models along with their type, context-window size, and maximum output length.



\begin{table*}[t]
\small
\centering
\caption{Evaluated LLMs ordered by variant type, context-window capacity, and max output length.}
\label{tab:llms}
\begin{tabular}{llcrr}
\toprule
\textbf{Model} & \textbf{Model ID} & \textbf{Type} & \textbf{Context Window} & \textbf{Max Output} \\
                    &                     &                    & (tokens) & (tokens)  \\
\midrule
Grok 4 Fast Non Reasoning \cite{xai}        & \texttt{grok-4-non-reasoning}                 & General-purpose            & 2M & N/A  \\
GPT-4.1 \cite{openai_docs}           & \texttt{gpt-4.1}                 & General-purpose                            & 1M   & 32K  \\
Llama 4 Scout \cite{openrouterLlamaScout}     & \texttt{llama-4-scout}           & General-purpose                & 328K  & 16K  \\
Claude Sonnet 4 \cite{anthropic_docs}  & \texttt{claude-sonnet-4}       & General-purpose & 200K & 64K  \\
DeepSeek V3.1 Chat \cite{DeepSeek_Docs}      & \texttt{deepseek-chat}             &  General-purpose                           & 128K  & 8K  \\
Grok 4 Fast Reasoning \cite{xai}    & \texttt{grok-4-reasoning}             & Reasoning-optimized & 2M & N/A  \\
Gemini 2.5 Flash \cite{googleGemini}  & \texttt{gemini-2.5-flash}        & Reasoning-optimized             & 1M   & 65K  \\
Gemini 2.5 Pro \cite{googleGemini}   & \texttt{gemini-2.5-pro}          & Reasoning-optimized                & 1M   & 65K  \\
Llama 4 Maverick \cite{openrouterLlamaMaverick}  & \texttt{llama-4-maverick}        & Reasoning-optimized                & 1M   & 16K  \\
GPT-5 mini \cite{openai_docs}           & \texttt{gpt-5-mini}                 & Reasoning-optimized                & 400K & 128K \\
Claude Sonnet 4 Thinking \cite{anthropic_docs}  & \texttt{claude-sonnet-4-think} & Reasoning-optimized & 200K & 64K  \\
DeepSeek V3.1 Reasoner \cite{DeepSeek_Docs}      & \texttt{deepseek-reasoner}             & Reasoning-optimized                & 128K  & 64K  \\
\bottomrule
\end{tabular}

\vspace{1mm}
\footnotesize{Notes: ``K'' = thousand tokens, ``M'' = million tokens, ``N/A'' = not specified.}
\end{table*}

\subsection{Blueprints and Dataset}
\label{sec:datasets}

To evaluate LLM-based analysis in a controlled and reproducible setting, we derive \emph{blueprints} from existing UVL feature models rather than from raw scoping inputs. While scoping artifacts represent the natural source for early validation, they are rarely standardized and lack solver-based ground truth. In contrast, UVL models are publicly available, semantically precise, and compatible with solver analysis, enabling reproducible comparison against an oracle.

We collected models from two established repositories, \textsc{UVLHub}~\cite{Romero2024} and the \emph{Feature-Model-Benchmark v1.0}~\cite{Sundermann2024paper}, and selected \textit{ten} representative cases covering both toy (e.g., \texttt{SW}, \texttt{SMW}) and large, real-world product lines (e.g., \texttt{BDB}, \texttt{CNNl/f}). Each model contains at least one cross-tree constraint and varies widely in feature count, tree depth, and constraint density.

For each UVL model, we constructed a corresponding \emph{blueprint} by restating its hierarchy and constraints in constrained natural language, as described in \textit{Section~\ref{sec:workflow}}. For \textit{generalization-related AOs}, each blueprint was paired with its \emph{variant} obtained by swapping selected relationships (mandatory $\leftrightarrow$ optional, or $\leftrightarrow$ alternative). This setting yields controlled pairs that allow testing if one blueprint’s variability space includes the other. 

All blueprints were manually verified for semantic equivalence to their UVL sources. \textit{Table~\ref{tab:fms}} summarizes structural metrics and blueprint sizes, ranging from fewer than 100 to over 70,000 tokens, thus spanning both compact and complex models.

\begin{table*}[t]
\small
\caption{Feature models used in the experiments with structural metrics and blueprint sizes.}
\label{tab:fms}
\centering 
\begin{tabular}{llrrrrrrrrrrrr}\toprule
\multicolumn{2}{l}{Feature Model}            & \texttt{SW} & \texttt{SMW} & \texttt{IDE} & \texttt{SMG} & \texttt{COM} & \texttt{SEA} & \texttt{CVE} & \texttt{BDB} & \texttt{CNNl} & \texttt{CNNf} \tabularnewline \midrule
\multicolumn{2}{l}{\#Features}               & 6  & 13  & 14  & 33  & 48  & 145 & 169 & 117 & 3,296 & 6,867  \tabularnewline  
\multicolumn{2}{l}{\#Relationships}          & 4  & 8   & 11  & 19  & 25  & 73  & 15  & 54  & 1,561 & 3,516 \tabularnewline 
\multicolumn{2}{l}{\#Cross-tree constraints} & 1  & 2   & 2   & 4   & 21  & 13  & 153 & 282 & 76   & 9 \tabularnewline 
\multicolumn{2}{l}{Tree Depth}                    & 2  & 2   & 2   & 3   & 3  & 10  & 4 & 5 & 10   & 11 \tabularnewline 
\midrule
\multicolumn{2}{l}{\#Blueprint tokens}       & 68 & 143 & 178 & 360 & 740 & 1,456 & 2,367 & 3,780 & 33,823 & 71,812 \tabularnewline
\bottomrule
\multicolumn{12}{l}{\footnotesize{Abbreviations: SW=Sandwich, SMW=Smartwatch, IDE=IDE product line, SMG=Strategy Mobile Game,}} \\
\multicolumn{12}{l}{\footnotesize{COM=Computer, SEA=Subsea Control System, CVE=Cybersecurity Vulnerability, BDB=Berkeley DB,}} \\
\multicolumn{12}{l}{\footnotesize{CNNl=light CNN architectures, CNNf=full CNN architectures.}} \\
\multicolumn{12}{l}{\footnotesize{Metrics extracted with FLAMA~2.0.1 and Glucose3; blueprint token counts measured in tokens.}}\\ 
\end{tabular}
\end{table*}

\subsection{Analysis Operations}

We consider 16 commonly used AOs from feature model research and practice~\cite{Benavidesetal2013,Galindo2019}, covering both structural metrics (e.g., feature counts, tree depth) and solver-based reasoning tasks (e.g., satisfiability, core features, configuration counting). \textit{Table~\ref{tab:fmao}} lists all evaluated AOs. Each AO is implemented through a dedicated prompt template that defines its reasoning steps and enforces a standardized output format for comparison with the solver-based oracle. \texttt{AO12} (\#valid configurations) is limited to the first eight FMs due to scalability constraints on the largest cases (\texttt{CNNl}, \texttt{CNNf}). For \texttt{AO16} (\emph{generalization}), each test involves a pair of blueprints: the original and its modified variant (see \textit{Section~\ref{sec:datasets}}).


\begin{table}[t]
\small
\centering 
\caption{AOs considered in the experiments.}
\label{tab:fmao}
\begin{tabular}{cl|cl}
\toprule
ID & Solver-free AOs & ID & Solver-based AOs \\
\midrule
\texttt{AO1}  & \#Features              & \texttt{AO10} & Satisfiable/Void \\
\texttt{AO2}  & \#Leaf Features         & \texttt{AO11} & Configuration Satisfiable \\
\texttt{AO3}  & Tree Depth              & \texttt{AO12} & \#Valid Configurations \\
\texttt{AO4}  & \#Mandatory Features    & \texttt{AO13} & Core Features \\
\texttt{AO5}  & \#Optional Features     & \texttt{AO14} & Dead Features \\
\texttt{AO6}  & \#Or Groups             & \texttt{AO15} & False Optional Features \\
\texttt{AO7}  & \#Alternative Groups    & \texttt{AO16} & Generalization \\
\texttt{AO8}  & \#Requires              &      & \\
\texttt{AO9}  & \#Excludes              &      & \\
\bottomrule
\end{tabular}
\end{table}

\subsection{Evaluation Protocol}
\label{sec:evaluation_protocol}

\begin{figure*}[t]
    \centering
    \includegraphics[width=\textwidth]{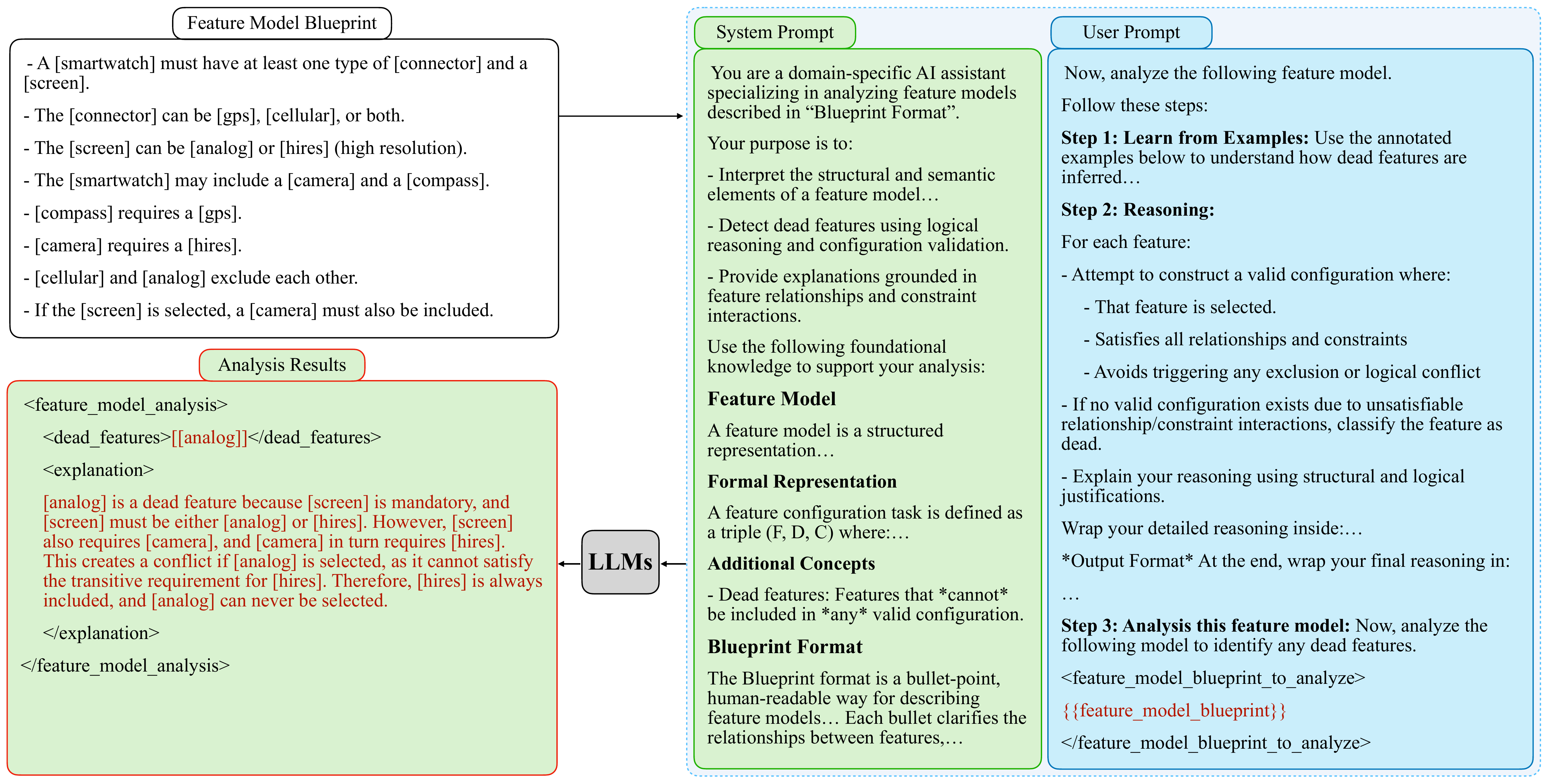}
    \caption{Prompting \emph{pipeline} for \emph{dead feature detection}, showing how the system and user prompts guide reasoning and how the LLM outputs XML results for solver comparison.}
    \label{fig:prompting-pipeline}
\end{figure*}

\paragraph{Prompt Design}
For each AO, we use a three-part pipeline illustrated in \textit{Figure~\ref{fig:prompting-pipeline}} using dead feature detection as an example. The figure shows how the \textit{system prompt}, \textit{user prompt}, and \textit{output contract} work together to guide the LLM through the analysis of a smartwatch blueprint. We now describe each component:

\begin{itemize}
    \item \textbf{System prompt.} The system prompt defines the model’s role as a domain-specific assistant for feature model analysis. It introduces the blueprint format, explains the elements of feature models (features, hierarchy, and cross-tree constraints), and specifies the semantics and expected outcome of the target AO using precise definitions and result formats.


    \item \textbf{User prompt.} Combines task instructions, illustrative examples, and the target input. It begins with a \emph{Learn from Examples} section containing 2–4 exemplars that pair blueprints with their corresponding AO results. A subsequent \emph{step-by-step procedure} describes the reasoning strategy (e.g., feature enumeration, constraint propagation, contradiction checking), followed by the blueprint to be analyzed. This design grounds the task in concrete examples and guides the model through a consistent, reproducible reasoning process.

    \item \textbf{Output contract.} Each AO uses an XML-based output schema to ensure machine-checkable results. The contract enforces standardized tags (e.g., \texttt{<dead\_features>}, \texttt{<core\_features>}, \texttt{<feature\_model\_analysis>}) with canonical content: integers for counts, \texttt{true}/\texttt{false} for booleans, and newline-separated or tag-enclosed lists for feature sets. Any deviation (e.g., unparseable text, missing tags, malformed lists) is automatically considered incorrect. For set-valued AOs, exact set equality with the solver-based oracle is required. Models are also instructed to include concise natural language justifications, such as explaining why a feature is classified as dead.
\end{itemize}


This uniform pipeline ensures consistent results across models and AOs by prioritizing fair capability comparison over tailored optimization. Full templates for all 16 AOs are provided in the replication package\footnote{
  Replication package: \url{https://github.com/AIG-ist-tugraz/llm-blueprint-analysis}
}.



\paragraph{Inference.}
Each triple (including model, blueprint, and AO) is executed once with deterministic decoding (\texttt{temperature=0}, greedy) via the provider’s public API only, i.e., no tool use, no solver calls, no post-processing. We record request/response payloads, the raw XML, the rationale, and wall-clock runtime. Besides, timeouts and truncations are kept (and evaluated) as produced.

\paragraph{Metrics and error taxonomy.}
We report the results along the following two metrics:

\begin{itemize}
\item \textbf{Accuracy.} Exact-match agreement with the FLAMA oracle, broken down by AO, blueprint, and model family (general-purpose vs.\ reasoning-optimized).
\item \textbf{Cost.} End-to-end \emph{runtime} (in seconds) and \emph{token usage} \allowbreak(prompt{+}completion).
\end{itemize}

Errors are categorized into four mutually exclusive \emph{failure modes} aligned with our analysis: (i) \emph{Unparseable} (violates XML contract), (ii) \emph{Format-correct but wrong} (includes \emph{semantic slips} such as misreading alternative vs.\ mandatory), (iii) \emph{Partial/Truncated} (incomplete outputs due to context/output limits), and (iv) \emph{Hallucinated elements} (items not in the blueprint).

\subsection{Baseline Oracle and Implementation}

We use FLAMA~2.0.1~\cite{Galindo2023flama} as the solver-based oracle, executing AOs on UVL inputs through the Glucose3 SAT solver and the DD library to obtain exact and reproducible ground-truth results for all comparisons. The LLM-based analysis engine and the evaluation framework are implemented in Python~3.10 within a unified evaluation harness. To ensure modularity and reproducibility, we employ LangChain~\cite{langchain} for standardized API integration, and LangGraph~\cite{langgraph} for orchestrating prompting pipelines. All prompts, blueprints, model settings, evaluation scripts, raw outputs, and per-run runtimes are publicly released in the replication package\footnotemark[1].




\section{Results}
\label{sec:results}

Full tables and results are shown in Appendix\footnote{
  Supplementary appendix: \url{https://doi.org/10.5281/zenodo.17913681}
}.



\subsection{RQ1: Accuracy of LLM-based AOs}
\label{sec:rq1-accuracy}

\begin{figure}[t]
    \centering
    \includegraphics[width=\linewidth]{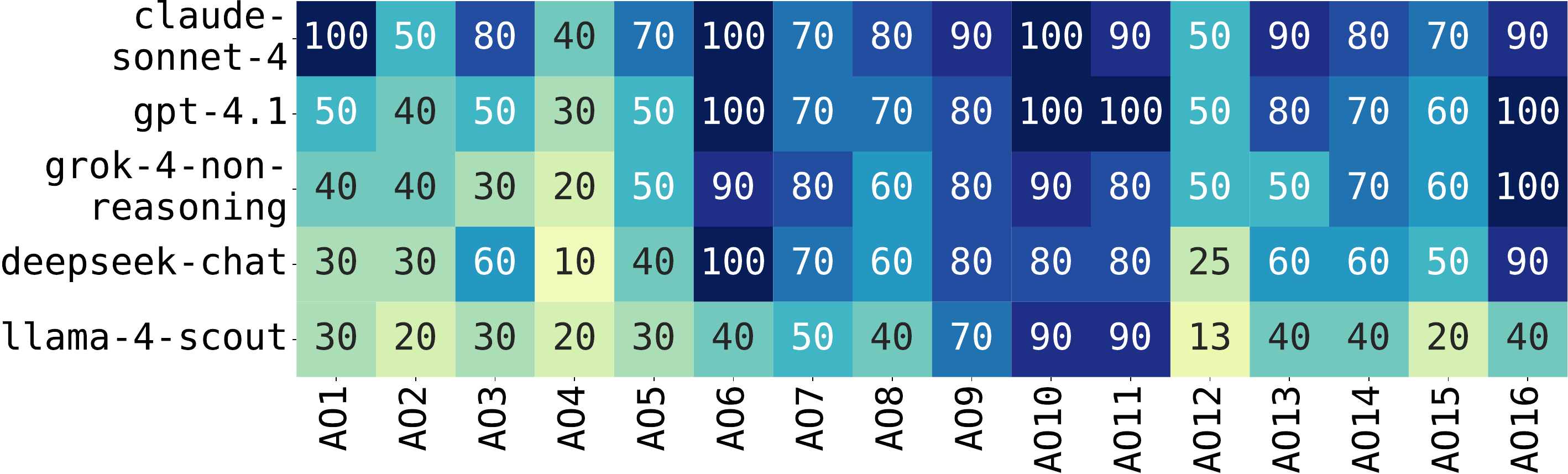}
    \caption{Accuracy (\%) of general-purpose LLMs across AOs.}
    \label{fig:heatmap_ao_general_all_aos}
\end{figure}

\begin{figure}[t]
    \centering
    \includegraphics[width=\linewidth]{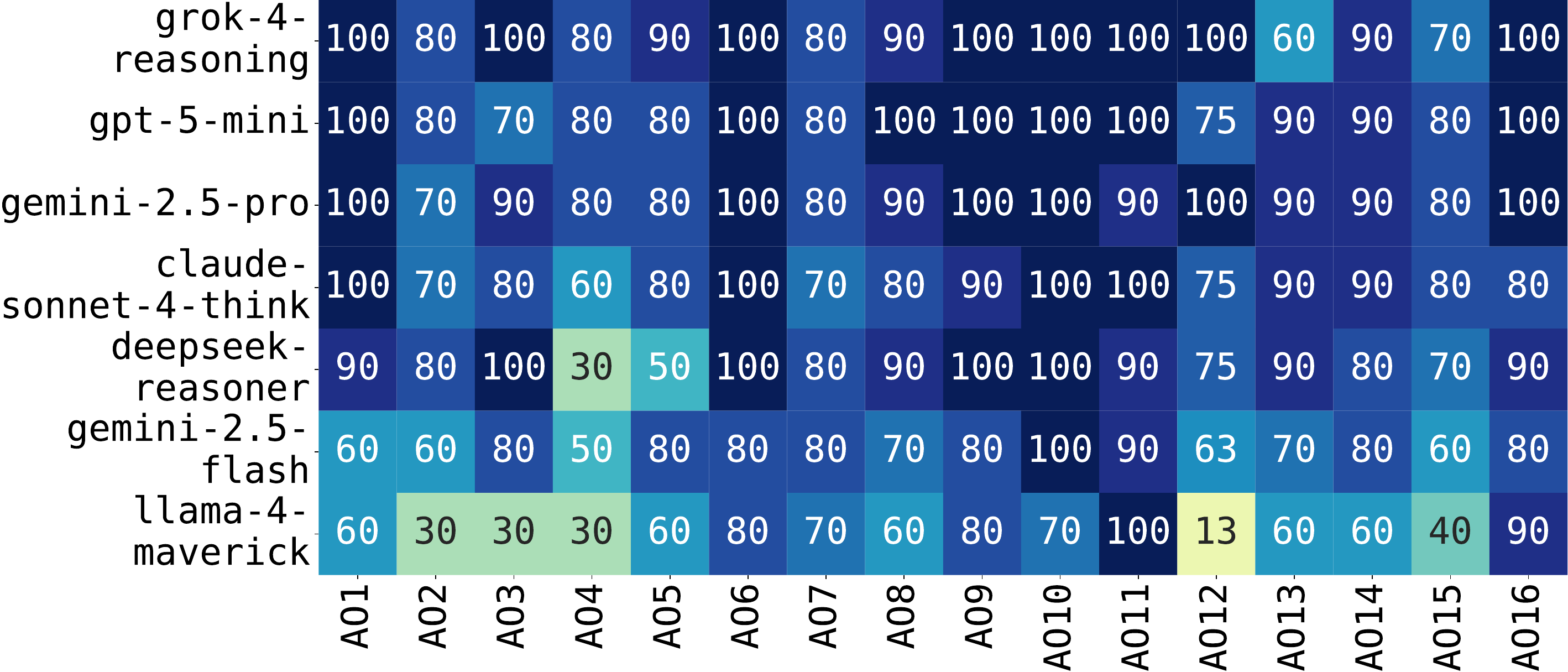}
    \caption{Accuracy (\%) of reasoning-opt. LLMs across AOs.}
    \label{fig:heatmap_ao_reasoning_all_aos}
\end{figure}

\textit{Figure~\ref{fig:heatmap_ao_general_all_aos}} and \textit{Figure~\ref{fig:heatmap_ao_reasoning_all_aos}} summarize accuracy values across all 16 AOs. Overall, reasoning-optimized LLMs clearly outperform general-purpose ones, achieving an average accuracy of 81.1\% compared to 61.0\%. 
This gap reflects stronger multi-step reasoning and constraint handling. 
Accuracy patterns, however, vary markedly across operation types and blueprint complexity.


\paragraph{Accuracy Across AOs.}
Solver-based AOs yield higher accuracies than solver-free ones for both model families: 68.8\% vs.\ 56.0\% for general-purpose and 83.5\% vs.\ 79.2\% for reasoning-optimized LLMs. Rather than requiring exhaustive enumeration, these solver-based tasks are \emph{operationalized as verification}. For instance, in \texttt{AO10}, the model is asked to \emph{search for contradictions} in the blueprint and concludes ``satisfiable'' if none are found. \texttt{AO11} analogously checks whether a given configuration \emph{violates any constraint} before deciding its satisfiability. \texttt{AO16} reduces to judging whether one model \emph{subsumes} another (i.e., no counterexample is identified). This verification-style framing lowers the need for complete search and tends to be more robust to local parsing noise. By contrast, solver-free AOs (e.g., \texttt{AO4--AO9}) require the model to \emph{parse and count} structural relationships such as \emph{mandatory}, \emph{optional}, or \emph{alternative} features; small semantic misunderstandings, e.g., interpreting ``\textit{A must have B or C}'' as two mandatory children instead of an alternative group, lead to systematic counting errors and lower accuracies.

Difficulty levels vary markedly across AOs and model families. For general-purpose LLMs, the hardest AOs are \texttt{AO2} (leaf count), \texttt{AO4} (mandatory count), and \texttt{AO12} (\#valid configurations), all averaging below 50\% accuracy, reflecting persistent semantic errors and limited reasoning capacity. In contrast, reasoning-optimized LLMs achieve high performance ($\geq$85\%) on most tasks but still struggle with \texttt{AO4} (58.6\%) due to semantic misunderstandings, and moderately on \texttt{AO12} and \texttt{AO15}, which demand constraint propagation or enumeration. Across both families, the easiest AOs are \texttt{AO6}, \texttt{AO9}, \texttt{AO10}, \texttt{AO11}, and \texttt{AO16} (all above 85\%), which involve simpler verification steps rather than complex structural reasoning.

\begin{figure}[t]
    \centering
    \includegraphics[scale=0.25]{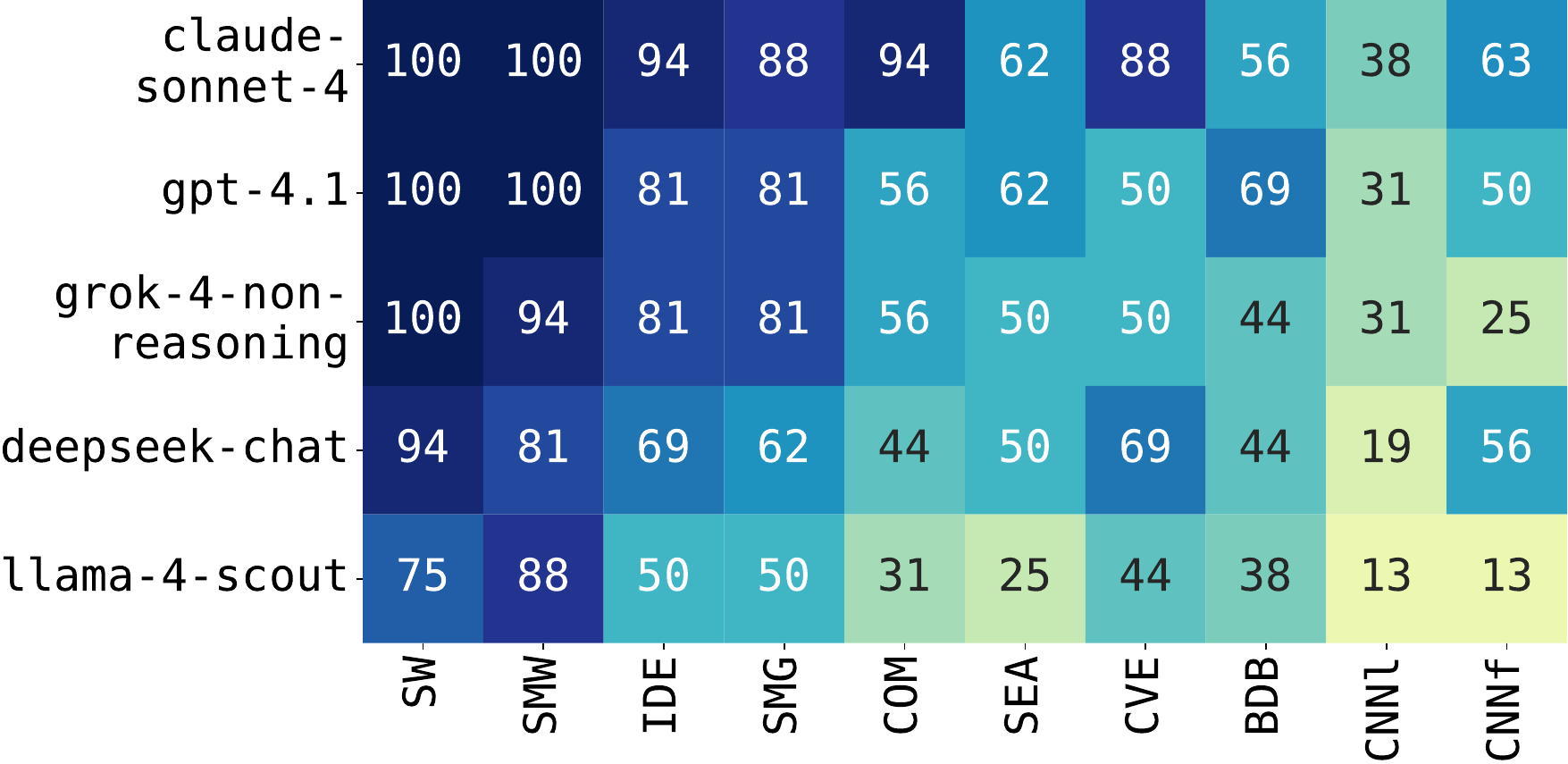}
    \caption{Accuracy (\%) of general-purpose LLMs on 16 AOs across 10 blueprints.}
    \label{fig:heatmap_bp_general_all_aos}
\end{figure}

\begin{figure}[t]
    \centering
    \includegraphics[scale=0.25]{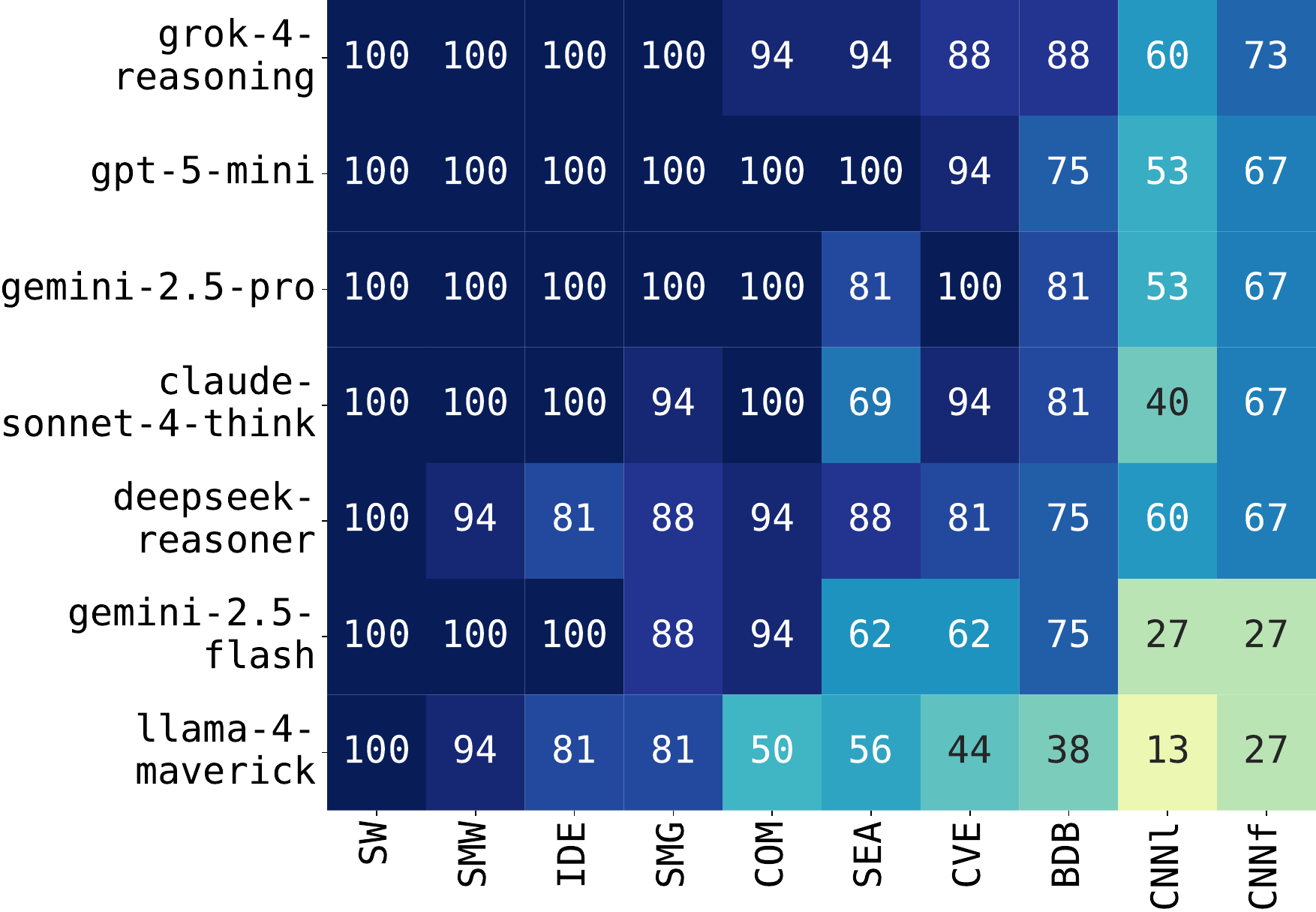}
    \caption{Accuracy (\%) of reasoning-optimized LLMs on 16 AOs across 10 blueprints.}
    \label{fig:heatmap_bp_reasoning_all_aos}
\end{figure}

\paragraph{Accuracy Across Blueprints.}
\textit{Figures~\ref{fig:heatmap_bp_general_all_aos} \& ~\ref{fig:heatmap_bp_reasoning_all_aos}} show accuracies aggregated by blueprint. Across both model families, accuracy decreases steadily with increasing blueprint complexity. All models achieve near-perfect performance on small and shallow blueprints (\texttt{SW}, \texttt{SMW}), moderate accuracy on medium-sized ones (\texttt{IDE}, \texttt{SMG}, \texttt{COM}), and substantial drops on complex models such as \texttt{SEA}, \texttt{BDB}, \texttt{CNNl}, and \texttt{CNNf}. These challenging cases combine large feature counts (up to 7,000), deep hierarchies (depth $\geq$10), and numerous cross-tree constraints ($>$300), which amplify semantic misunderstandings and reasoning limitations. Importantly, no single model achieves 100\% accuracy across all blueprints, highlighting the difficulty of handling large, constraint-dense feature models.

\paragraph{Model Ranking.}
Among reasoning-optimized LLMs, three models form the top-performing group: Grok 4 Fast Reasoning (89.7\%), GPT-5 mini (88.9\%), and Gemini 2.5 Pro (88.2\%) (see Figure~\ref{fig:heatmap_bp_reasoning_all_aos}). They achieve perfect scores on several simple AOs and maintain stable performance on complex ones, suggesting potential for ensemble-based aggregation. The second-tier reasoning models, Claude Sonnet 4 Thinking (84.5\%) and DeepSeek Reasoner (82.8\%), show moderate drops on \texttt{AO12} (both 75\%) and \texttt{AO15} (80\% and 70\%). They also exhibit structural counting weaknesses: Claude Sonnet 4 Thinking performs lower on \texttt{AO4} (60\%) and \texttt{AO2}/\texttt{AO7} (70\%), while DeepSeek Reasoner falls sharply on \texttt{AO4} (30\%) and \texttt{AO5} (50\%). Gemini 2.5 Flash and Llama 4 Maverick perform worse (60--75\%), mainly due to context overflow and output truncation. Within the general-purpose family (see \textit{Figure~\ref{fig:heatmap_bp_general_all_aos}}), Claude Sonnet 4 leads with 78.3\%, followed by GPT-4.1 and Grok 4 Fast Non Reasoning (60--68\%), while Llama 4 Scout is the weakest (42.7\%).

\paragraph{Takeaway.}
Reasoning-optimized models consistently outperform general-purpose ones on blueprint AOs, but accuracy declines with blueprint size/depth and remains uneven for structural counts (e.g., mandatory/alternative counting).

\subsection{RQ2: Efficiency of LLM-Based AOs}
\label{sec:rq2-efficiency}

Since AOs demand high precision, computational cost is meaningful only in relation to achieved accuracy. A model that spends more time but delivers correct results is preferable over a faster yet unreliable one. \textit{Table~\ref{tab:costs}} reports average runtime, token usage, and accuracy across all AOs and blueprints. Reasoning-optimized LLMs are notably more resource-intensive, requiring 2.4k--9.5k~seconds and 500K--840K tokens, compared to 0.7k--1.9k~seconds and 370K--440K tokens for general-purpose models. The symbolic solver baseline (FLAMA) remains by far the fastest (18.45~seconds) but cannot process semi-formal blueprints directly.

\begin{table}[t]
\small
\centering
\caption{
Average runtime (in second), token usage, and accuracy across all AOs and blueprints for representative models.
}
\label{tab:costs}
\begin{tabular}{lccc}
\toprule
\textbf{Model ID} & \textbf{Runtime (s)} & \textbf{Tokens} & \textbf{Accuracy (\%)} \\
\midrule
\texttt{grok-4-reasoning}     & 3,046.7 & 505,874 & 89.7 \\
\texttt{gemini-2.5-pro}        & 2,390.8 & 602,251 & 88.2 \\
\texttt{gpt-5-mini}            & 6,118.3 & 843,251 & 88.9 \\
\texttt{deepseek-reasoner}     & 9,508.4 & 561,290 & 82.8 \\
\texttt{claude-sonnet-4}       & 1,168.5 & 439,523 & 78.3 \\
\texttt{llama-4-scout}         & 680.2   & 374,608 & 42.7 \\
\midrule
FLAMA (Solver)$^{\dagger}$        & 18.45 & -- & 100.0 \\
\bottomrule
\end{tabular}

\vspace{1mm}
{\footnotesize{$^{\dagger}$Results on formal inputs; solver cannot process semi-formal blueprints.}}
\end{table}

\paragraph{Accuracy–Cost Balance.}
High computational cost does not necessarily guarantee better accuracy. For instance, DeepSeek Reasoner is the slowest model (9,508~seconds) yet achieves only 82.8\% accuracy, while Grok 4 Fast Reasoning reaches 89.7\% with one-third of the runtime. In contrast, Gemini 2.5 Pro delivers similar accuracy (88.2\%) with the lowest runtime among reasoning models (2,391~seconds), representing the most efficient high-accuracy trade-off. GPT-5 mini attains top accuracy (88.9\%) but at substantially higher runtime and token cost, suggesting diminishing returns. Meanwhile, general-purpose models such as Claude Sonnet 4 offer a moderate balance (78.3\%, 1,169~seconds), and lightweight models like Llama 4 Scout are fast yet unreliable (42.7\%).


\subsection{RQ3: Error Analysis}
\label{sec:error-analysis}

We identify three dominant failure modes. First, \emph{semantic slips} in solver-free AOs (\texttt{AO4--AO9}), most notably misreading ``A must have B or C'' as two mandatory children instead of an \emph{alternative} group. Second, \emph{incomplete propagation or enumeration} in solver-based AOs, especially \texttt{AO12} (\#valid configurations) and \texttt{AO15} (\#false optional), which require constraint reasoning beyond surface parsing. Third, \emph{context and output limits} on very large blueprints (\texttt{CNNl}, \texttt{CNNf}), causing truncation or early stopping. These problems intensify in deep or constraint-dense models (\texttt{SEA}, \texttt{BDB}, \texttt{CNNl/f}) and are most pronounced in \texttt{AO12}, \texttt{AO15}, and structural AOs (\texttt{AO4--AO9}). Detailed counts per model and AO are in Appendix\footnotemark[2].


Model behaviors reflect these patterns. Grok 4 Fast Reasoning, GPT-5 Mini, and Gemini 2.5 Pro make few but systematic propagation errors. Claude Sonnet 4 often overcounts mandatory features due to confusion between or/alternative and mandatory relationships. DeepSeek Reasoner shows partial outputs while Gemini 2.5 Flash and Llama 4 Maverick suffer truncation near context limits. Lightweight models (Llama 4 Scout, DeepSeek Chat) tend to produce format-correct but semantically wrong answers. These recurring patterns underscore the need for disambiguation rules, output chunking, and model ensembles.

\subsection{Overall Synthesis}

Across RQ1–RQ3, reasoning-optimized LLMs approach solver-level accuracy on many analyses but remain sensitive to structural ambiguity, scale, and context limits. Most errors stem from incomplete constraint reasoning rather than random noise. In summary, LLMs can already serve as reliable early validators of variability models when paired with simple safeguards and ensemble strategies.

\section{Discussion}

\label{sec:discussion}

\paragraph{Where LLMs fit in early scoping.}
Our results support a pragmatic placement of LLMs at the \emph{earliest} stage, when blueprints are semi-formal and solver-ready model is not yet available. In this context, reasoning-oriented models provide \emph{verification-style} feedback with near-solver accuracy (e.g., for \texttt{AO10}, \texttt{AO11}, \texttt{AO16}), quick structural sanity checks (e.g., for \texttt{AO1}, \texttt{AO3}, \texttt{AO6}, \texttt{AO9}), and concise explanations that help stakeholders refine scope decisions before formalization.

\paragraph{Accuracy and explainability over speed.}
Because early validation is iterative and not latency-critical, we prioritize correctness and interpretability over raw speed. The observed runtime/token overheads are acceptable in practice given the benefit of solver-like guidance \emph{without} requiring a formal model. Reasoning-optimized models (e.g., Grok 4 Fast Reasoning, Gemini 2.5 Pro, GPT-5 mini) therefore offer a practical operating point: near-solver accuracy with stable formatting and useful rationales that fit within minutes-scale review cycles typical of scoping workshops.

\paragraph{Leveraging model complementarity.}
Top performers exhibit complementary strengths across blueprints and AOs. Grok 4 Fast Reasoning demonstrates exceptional robustness on large, constraint-dense models such as \texttt{BDB}, \texttt{CNNl}, and \texttt{CNNf}, where other models suffer truncation or incomplete propagation. In contrast, GPT-5 mini achieves the highest consistency on mid-size blueprints (\texttt{COM}, \texttt{SEA}) and complex solver-based tasks (\texttt{AO12}, \texttt{AO15}), benefiting from deeper constraint reasoning. Gemini 2.5 Pro performs best on compact to medium blueprints (\texttt{CVE}, \texttt{SMG}) and excels in maintaining strict XML conformance and stable formatting even under long outputs. These differences suggest that the models compensate for each other’s weaknesses: Grok handles scale and constraint density, GPT-5 mini excels in fine-grained constraint reasoning, and Gemini maintains strict output conformance. Simple ensemble aggregation (majority vote or confidence-weighted fusion) mitigates idiosyncratic errors and raises reliability with minimal engineering, matching the “low-friction” goal of early validation.


\paragraph{Actionable guidance.}
To harden early analyses, we recommend four concrete measures aligned with the observed failure modes:
(i) add short disambiguation rules that test group semantics before any counting (distinguish \emph{or}/\emph{alternative} from \emph{mandatory}) to reduce semantic slips in \texttt{AO4--AO9}, (2) plan outputs for long blueprints by chunking lists and capping free text to avoid truncation and partial results on large cases (e.g., \texttt{CNNl}/\texttt{CNNf}), (iii) use model ensembles on medium and large blueprints (majority vote or confidence-weighted fusion across top models) to smooth per-model weaknesses in propagation and enumeration (notably \texttt{AO12} and \texttt{AO15}), (iv) maintain human oversight through stakeholder review and verification protocols to prevent automation bias, as approximately 1 in 10 analyses may contain errors.


\paragraph{Threats to Validity}
Our evaluation uses blueprints derived from UVL models to ensure reproducibility and oracle comparability. While this enables consistent benchmarking, it may under-represent ambiguity and noise found in truly informal scoping artifacts. Moreover, the 16 evaluated AOs cover representative but not exhaustive FM tasks. We also evaluated a broad set of public LLM APIs, but not all commercially or academically available models. Results may vary for future versions or domain-specific fine-tuning. Finally, our strict exact-match scoring penalizes partially correct outputs, which may underestimate the practical reasoning capabilities of models in near-missing cases.

\section{Conclusion and Future Work}
We evaluated whether off-the-shelf LLMs can perform feature model AOs directly on semi-formal blueprints, where early feedback is most valuable. Reasoning-optimized models reach near-solver accuracy and offer actionable explanations, making them practical assistants for SPL scoping. Their minute-scale runtime is acceptable for accurate, automation-ready feedback before formalization. Remaining issues, such as semantic slips and incomplete propagation, can be mitigated through ensemble use and prompt safeguards. Overall, LLMs provide a feasible path to automated, explainable variability checks prior to solver-based verification, though human oversight remains necessary given their 88--89\% accuracy.


Future work will expand both the analytical scope and the workflow.
We plan to include additional AO (e.g., redundancy, atomic sets, diagnosis/conflict sets) to better approximate the full AO catalogue, and to conduct \emph{workflow evaluations in practice} with domain engineers to assess usability and feedback quality.
We also aim to support \emph{blueprint-to-UVL translation} with round-trip consistency checks and to develop \emph{learned routers} that select the most reliable model.
Finally, we envision extending the workflow toward \emph{scope completeness checks}, where LLMs ensure traceability between goals, scenarios, and features—detecting orphaned or missing links to reduce underscoping and misalignment with business objectives.

\begin{acks}
The work presented in this paper has been developed within the research project \textsc{GenRE} funded by the Austrian Research Promotion Agency under the project number $915086$.
\end{acks}

\printbibliography

\end{document}